# Fuzzy permutation time irreversibility for nonequilibrium analysis of complex system*

YAO Wenpo
School of Chemistry and Life Sciences, Nanjing University of Posts and Telecommunications, Nanjing 210023, China

## Abstract

Permutation time irreversibility is an important method to quantify nonequilibrium characteristics of complex systems; however, ordinal pattern is a coarse-graining alternative of temporal structure and cannot accurately represent detailed structural information. This study aims to propose a fuzzy permutation time irreversibility (fpTIR) by measuring the difference between vector elements based on a negative exponential function. The amplitude permutation of vector is constructed and its membership degree is calculated; then, the difference in probability distribution between the forward and backward sequences is measured for fpTIR. To compare and measure the system's complexity, the Shannon entropy is calculated as the average amount of information in the fuzzy permutation probability distribution, i.e., fuzzy permutation entropy (fPEn). According to the surrogate theory, mode series are generated using logistic, Henon, and first-order autoregressive systems to verify the fpTIR, which is then used to analyze the heartbeats of patients with congestive heart failure and healthy elderly and young participants from the PhysioNet database. Results suggest that the fpTIR effectively measures the system's nonequilibrium characteristics, thus improving the accuracy of heartbeat analysis. However, in analyzing probability distributions, the fpTIR and fPEn exhibit discrepancies in the chaotic series and even opposite results in the heartbeats, wherein the results of fpTIR are consistent with the theory of complexity loss in aging and disease. Overall, the fpTIR accurately characterizes the structure of the sequences and enhances the accuracy of the nonequilibrium analysis of complex systems, providing a theoretical basis for exploring complex systems from the perspectives of nonequilibrium dynamics and entropy complexity.

---

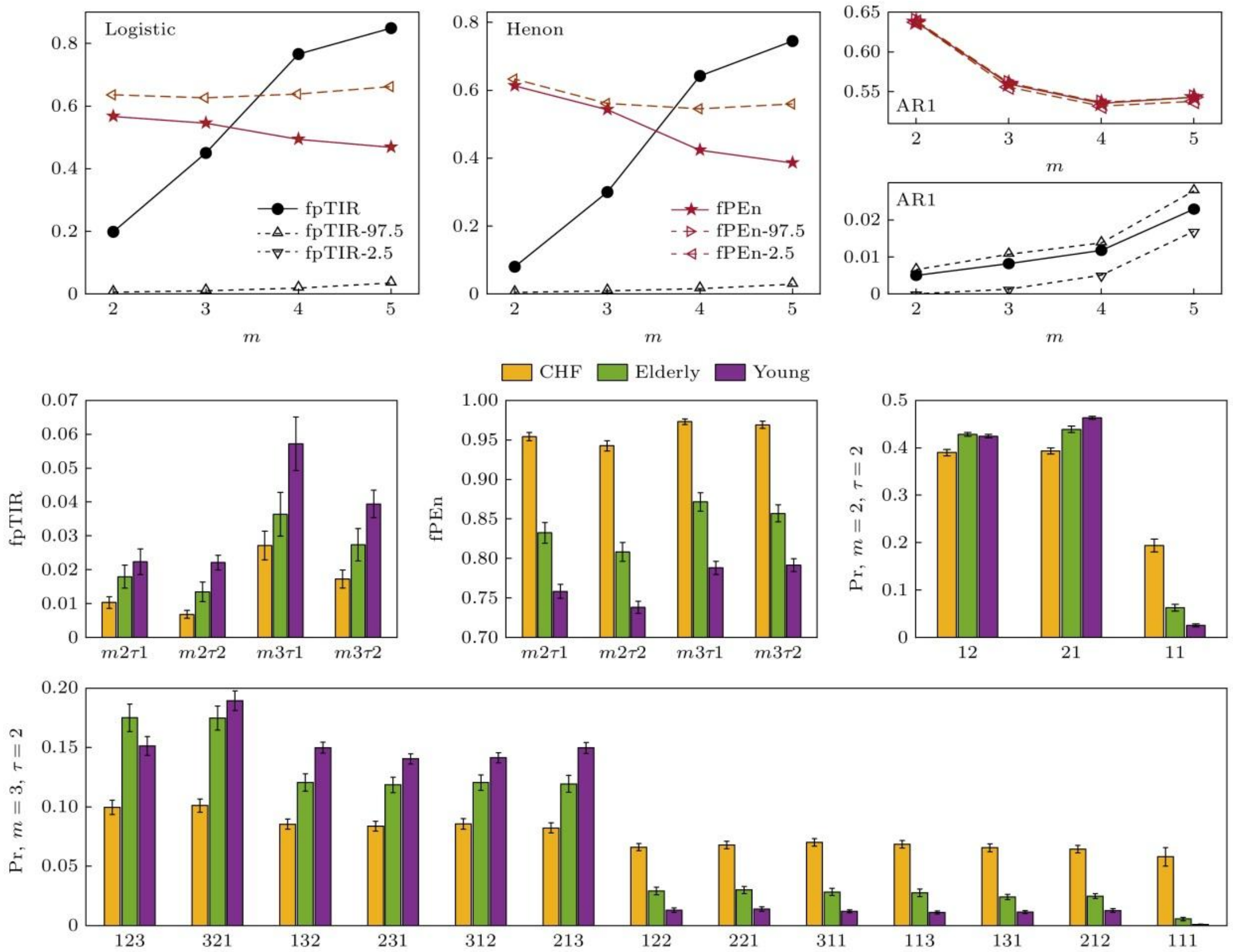

Keywords: fuzzy permutation; time irreversibility; permutation entropy; complex system; symbolic dynamics

**PACS:** 05.45.Tp; 05.70.Ln; 07.05.Mh; 02.70.Wz

## 1. Introduction

Time reversibility refers to the property of a system whose characteristics remain unchanged with the reverse order of time, otherwise it is time irreversibility (TIR)[1]. TIR is a property to measure the non-equilibrium system, and its quantification involves the calculation of the joint probability difference between forward and backward sequences or symmetric vectors of a sequence. Currently, quantitative TIR mainly focuses on the symbolic time series[2]. Because the permutation[3] effectively represents the structural characteristics of the sequence and does not need model setting, it has attracted wide attention in TIR analysis[4–9].

Martinez et al.[4] used the probability difference of permutation to quantify the TIR and revealed the non-equilibrium characteristics of complex systems, such as economy, meteorology, and physiology. Zanin et al.[5,6] verified the advantage of the probabilistic difference of forward-backward permutation sequences in complex economic signals. Yao et al. [1,7–9] studied the factors of permutation time irreversibility (pTIR) that influence complex system analysis, focusing on the forbidden permutation and equal values. These studies advanced the understanding and practical application of the theory of irreversibility of permutation time.

As a coarse-graining expression of the sequence structure, permutation not only simplifies the spatial characteristics, but also affects the results of TIR. First, temporal structures represented by the two basic ordinal patterns, namely the original permutation and amplitude permutation, are distinct[9], in which the amplitude permutation is the direct representation of the spatial structure. In time-irreversible quantization, especially that based on the probability difference of symmetric simplified vectors, the original permutation may bring bias[8]. Second, forbidden permutation is a disadvantageous factor for pTIR calculation[1]. Amigo et al.[10–12] studied the characteristics of forbidden permutation and proposed various nonlinear analysis methods combined with practical applications. Subsequently, the system information related to forbidden permutation from multiple perspectives, such as chaotic dynamics and correlation, have been studied[13,14].

However, in pTIR, forbidden permutations leading to division-based parameters (such as Kullback–Leibler distance, i.e., relative entropy) are not suitable for the calculation of probability differences of permutations[1,7–9,15]. In addition, equal values, which are usually ignored in permutation analysis, play a key role in pTIR. Equal values are key factors in establishing a comprehensive permutation structure[9], affect the probability distribution and permutation, and even cause different results[1,16,17]. They might generate a self-symmetric vector, which has a clear physical meaning in the analysis of time irreversibility, i.e., time reversibility and time symmetry[1,7–9]. To improve the permutation with equal values, Bian et al.[18] proposed a modified permutation by mapping the equal value onto the same symbol (rank), following which, Zunino[19] and Cuesta et al.[20] demonstrated the influence of equal values on complexity quantification through permutation entropy. Unfortunately, because the permutation is a coarse-grained representation of the vector space structure, it cannot reflect the absolute positional relationship between elements. Hence, the current optimization of permutation analysis cannot solve its lack of accuracy.

To achieve an accurate characterization of vector structure characteristics and improve the accuracy of time irreversibility analysis, this study proposes a fuzzy permutation time irreversibility (fpTIR), which transforms the difference between elements into the membership degree of permutation type. The paper is outlined as follows. First, the concept of time irreversibility and establishment of amplitude permutation are introduced. Then, the vector permutation membership is calculated, and the fpTIR is described in detail. Next, the model sequence and its surrogate data are constructed to verify the fpTIR. Lastly, fpTIR is compared with the fuzzy permutation entropy (fPEn) to study the complex characteristics of heartbeats in patients with congestive heart failure (CHF) and healthy young and elderly participants in PhysioNet database.

## 2. Theory

### *2.1 Time-irreversibility concept*

In statistics, if the system is invariant under the time-inverse scale, it is time reversible; otherwise, it is time irreversible. According to Weiss[21], a stationary process $X(t)$ is time reversible if the $\{X(t_1), X(t_2), \cdots, X(t_m)\}$ and $\{X(-t_1), X(-t_2), \cdots, X(-t_m)\}$ have the same joint probability distribution for all $t_1, t_2, \cdots, t_m$ and $m$. Based on works of Kelly[22], for all $n$ and $m$, if $\{X(t_1), X(t_2), \cdots, X(t_m)\}$ and $\{X(-t_1 + n), X(-t_2 + n), \cdots, X(-t_m + n)\}$ have the same joint probability distribution, the process is time reversible. In particular, $\{X(t_1), X(t_2), \cdots, X(t_m)\}$ and its symmetric $\{X(t_m), \cdots, X(t_2), X(t_1)\}$ have the same probability distribution when $n = t_1 + t_m$.

### *2.2 Fuzzy amplitude permutation*

The original permutation (OrP) and amplitude permutation (AmP) are two basic ordinal patterns[9], where OrP refers to the indexes of reorganized values in the original series, and AmP refers to the positions of the original values in the reordered series. OrP can locate the element position after sequence sorting and is widely used in scientific research and numerical analysis software, such as MATLAB's 'sort', PYTHON's 'argsort', and R's 'order' algorithms. AmP is a direct representation of the sequence spatial structure, and its use can avoid the bias that may exist in the permutation time irreversibility[1,9].

Given a vector $X(i) = \{x(i_1), x(i_2), \cdots, x(i_n), \cdots, x(i_L)\}$ of length $L$ and reordering its elements ascendingly to $X(j) = \{x(j_1), x(j_2), \cdots, x(j_n), \cdots, x(j_L)\}$, where $x(j_1) < x(j_2) \cdots < x(j_n) < \cdots < x(j_L)$, the AmP is constructed for the positions of original vector elements $x(i_n)$ in the ordered vector $X(j)$, i.e $\pi_j = (j_{1n}, j_{2n}, \cdots, j_{in}, \cdots, j_{Ln})$.

Equal values are the key to establishing permutation, and their existence may produce time-reversible or time-symmetric self-symmetric vectors, which affect permutation analysis[1,7–9].

To characterize equal values, their indexes can be modified to be the same as those in their corresponding groups[9]. First, the equal values are arranged according to their orders of occurrence, such as $x(j_1) = x(j_2) < \cdots < x(j_3) = x(j_4) = x(j_5)$. Then, the coordinates of the equalities are modified to the same value of the same group, such as the minimum value $x(j_1) = x(j_1) < \cdots < x(j_3) = x(j_3) = x(j_3)$, as suggested by Bian[18], or the maximum value $x(j_2) = x(j_2) < \cdots < x(j_5) = x(j_5) = x(j_5)$.

Permutation is a coarse-grained representation of the vector structure and cannot reflect the absolute difference between elements. To solve this problem, the standard vector is defined first. Then, the difference vector of the sorted vector is calculated. If its standard deviation is zero, it is the standard vector; otherwise, it is a strange vector. As the standard deviation of the difference vector increases, the specificity of the strange vector also increases; in other words, the difference between the difference and standard vectors increases.

To quantify the specificity of the strange vector, the membership degree is introduced into the AmP to construct the fuzzy permutation. For the vector $X(j)$ reorganized in ascending order, the difference of the adjacent elements is $x^{'}(j_i) = x(j_{i+1}) - x(j_i)$, which constructs the difference vector $X^{'}(j) = \{x^{'}(j_1), \cdots, x^{'}(j_i), \cdots, x^{'}(j_{m-1})\}$. The equal values in the $X(j)$ is merged to be an element, thus avoiding the influence on the calculation of the membership degree. According to the fuzzy theory[23], the negative exponential function[24,25] can be used to calculate the probability $p_j$ of the vector $X(j)$ belonging to the AmP, as shown in Equation (1) below:

$$p_j = f_{\mathrm{E}}(j) = \exp[-\alpha \cdot \sigma(j)]. \tag{1}$$

where $\alpha$ is a control parameter, and $\sigma(j)$ is the standard deviation of $X^{'}(j)$.

If the vector $X(j)$ contains only two elements, $\sigma(j)$ can be modified as the ratio of the standard deviation of the vector to the whole sequence. Otherwise, if $X(j)$ are all-equal elements, the membership $p_i$ is set to 1.

Fuzzy permutation is the combination of AmP and its membership $\pi_f = (\pi_j; p_j)$. Thus, the standard vector ascending sequence $\sigma(j) = 0$; that is, $p_j = 1$. If the strange vector $\sigma(j) > 0$, then $p_j < 1$, and $p_j$ decreases with increasing specificity.

Using the five-element vectors $X(1) = \{9,3,7,1,5\}$ and $X(2) = \{8.5,3.1,6.5,1.3,5.2\}$

as an example, the AmP is $\pi_j = (5,2,4,1,3)$ ; the ascending difference vectors are $X^{'}(1) = \{2,2,2,2\}$ and $X^{'}(2) = \{1.8,2.1,1.3,2\}$ ; and the standard deviations are $\sigma(1) = 0$ and $\sigma(2) = 0.35$. If the control parameter of (1) is $\alpha = 1$, then the membership degrees of $X(1)$ and $X(2)$ are $p_1 = 1$ and $p_2 = 0.7$, and their fuzzy permutations are $\pi_1 = (5,2,4,1,3; 1)$ and $\pi_2 = (5,2,4,1,3; 0.7)$, respectively.

*2.3 Time irreversibility based on fuzzy permutation*

For the sequence $X(t) = \{x(1), x(2), \cdots, x(t), \cdots, x(L)\}$, the vector $x_\tau^m(t) = \{x(t), x(t+\tau), \cdots, x(t+(m-1)\tau)\}$ of dimension $m$ and delay $\tau$ is reconstructed to obtain the vector sequence $X_\tau^m(t) = \{x_\tau^m(1), x_\tau^m(2), \cdots, x_\tau^m(t), \cdots, x_\tau^m(L-(m-1)\tau)\}$. Then, the fuzzy permutation $\pi_{\mathrm{f}}(t)$ is generated by the amplitude permutation and membership degree of the vector $x_\tau^m(t)$. Finally, the fuzzy permutation sequence is constructed as $\pi_\tau^m(t) = \{\pi_{\mathrm{f}}(1), \pi_{\mathrm{f}}(2), \cdots, \pi_{\mathrm{f}}(t), \cdots, \pi_{\mathrm{f}}(L-(m-1)\tau)\}$. The probability of fuzzy permutation $p(\pi)$ is the ratio between the accumulation of the amplitude permutation membership $p_j$ and the number of permutations, as shown in Equation (2). Time irreversibility is the probability difference of fuzzy permutation between the sequence $X(t)$ and its time reverse $X(-t)$. Considering the existence of forbidden permutation, $Y_s$, based on subtraction, is used to calculate the probability difference[1,7–9,15] of the fuzzy permutation, as shown in Equation (3), where $p_{\mathrm{f}}(\pi)$ and $p_{\mathrm{b}}(\pi)$ refer to the probability distribution of the fuzzy permutation of the forward and backward sequences, respectively. The procedure for calculating the fuzzy permutation irreversibility of a time series is shown in Fig. 1.

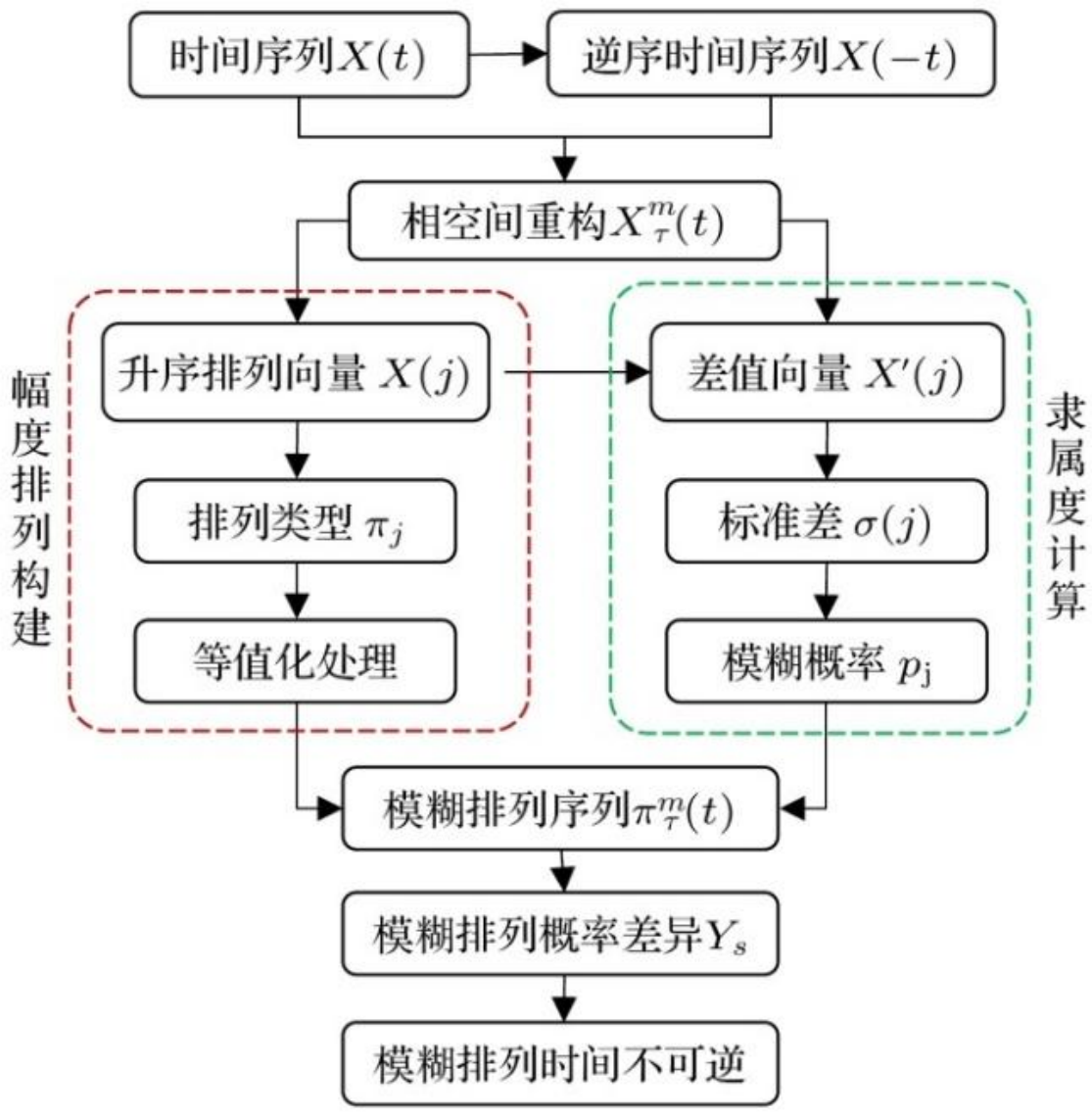

**Figure 1.** Flowchart of the calculation of fuzzy permutation time irreversibility.

$$p(\pi) = \frac{\sum p_j}{L-(m-1)\tau}, \quad (2)$$

$$\text{fpTIR} = \sum Y_s \langle p_{\text{f}}(\pi), p_{\text{b}}(\pi) \rangle = \sum p_{\text{f}}(\pi) \frac{p_{\text{f}}(\pi)-p_{\text{b}}(\pi)}{p_{\text{f}}(\pi)+p_{\text{b}}(\pi)}. \quad (3)$$

As a comparison, Shannon entropy is used to calculate the average information[26] of the probability distribution of the fuzzy permutation, herein, the (fuzzy permutation entropy fPEn), as shown in Equation (4). The fPEn is normalized by $\ln[k(m)]$, where $k(m)$ is the maximum number of permutation types[18] when the dimension is $m$; when the $m$ is 2, 3, 4 and 5, the $k(m)$ is 3, 13, 73 and 501, respectively.

$$\text{fPEn} = -\frac{\sum p(\pi) \ln p(\pi)}{\ln [k(m)]}. \quad (4)$$

## 3. Results

Model sequences are constructed to verify the effectiveness of fpTIR according to the surrogate theory, and real-world heartbeat data are analyzed.

### *3.1 Model-sequence verification*

First, chaotic series were constructed using logistic $(x_{t+1} = r \cdot x_t(1 - x_t),\ r = 4)$ and Henon $(x_{t+1} = y_t + 1 - \alpha x_t^2,\ y_{t+1} = \beta x_t,\ \alpha = 1.4$ and $\beta = 0.3)$ models, and a linear series was constructed based on a first-order autoregressive model (AR1, $x_{t+1} = \delta x_t + \xi_t$, where $\xi_t$ is Gaussian signal, $\delta = 0.3$). Then, 500 sets of surrogate data for each model series were constructed based on Schreiber's improved amplitude adjusted Fourier transform method[27,28].

According to the surrogate theory, if the fpTIR of the model series is between the 2.5% and 97.5% percentiles of the surrogate data, the model series and surrogate data are not significantly different; otherwise, if the fpTIR of the model series is greater than the 97.5% percentile or less than the 2.5% percentile of the surrogate data, the model series is significantly different from the surrogate data. The logistic, Henon, and AR1 model series and their surrogate data for the fpTIR and fPEn are shown in Fig. 2.

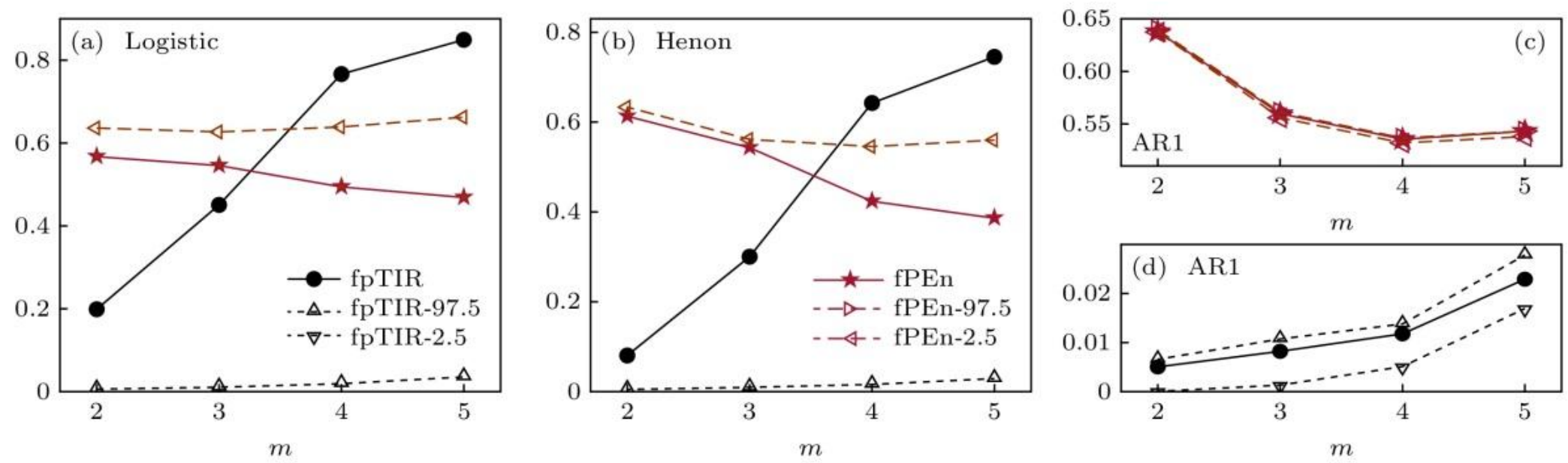

**Figure 2. The fpTIR and fPEn of the logistic, Henon, and AR1 series and their surrogates. The control parameter of the membership degree was $\alpha = 1$; the dimension and delay values were $m = 2$–$5$ and $\tau = 1$, respectively. The fpTIR-97.5, fpTIR-2.5, fPEn-97.5, and fPEn-2.5 denote the 97.5% and 2.5% of the fpTIR and fPEn of the surrogate data.**

As shown in Fig. 2, both the fpTIR and fPEn can effectively characterize the complex characteristics of the model sequence; however, they yielded different results in the chaotic sequence. In Fig. 2(a) and (b), the fpTIR of the logistic and Henon chaotic series was higher than the 97.5% percentiles of their surrogate data, whereas their fPEn is lower than the 2.5% percentiles. This discrepancy can be attributed to the fact that the fpTIR and fPEn have different analysis methods for the probability distribution of the fuzzy permutation. The surrogate data generated by iAAFT have higher randomness and unpredictability, and the probability distribution of its fuzzy permutation is more uniform. Therefore, the fpTIR of the chaotic series is significantly higher than that of the surrogate data, whereas its fPEn is significantly lower. Both the fpTIR and fPEn of the linear AR1 model are between the 2.5% and 97.5% percentiles of its surrogate data, as shown in Fig. 2(c) and (d). Therefore, according to the surrogate theory, the logistic and Henon series are nonlinear, whereas the AR1 model series are linear, which is consistent with the characteristics of the three model series.

### 3.2 FpTIR analysis of heartbeat signals

Human heartbeat is affected by various internal (such as nerves, body fluids, and hormones) and external (sound, light, temperature, and humidity) factors; thus, it is a highly dynamic and complex variable measure. The heartbeats of CHF patients and healthy elderly and young people in the PhysioNet database[29] are used to analyze the fpTIR. The data of 44 CHF patients [(55.5 ± 11.4) years old, aged 22–79 years] were collected from the chfdb and chfdb2 datasets[30], and those of 20 healthy elderly [(74.5 ± 4.4) years old, aged 68–85 years] and 20 healthy young [(25.8 ± 4.3) years old, aged 21–34 years] participants were collected from the Fantasia database[31]. The healthy elderly and young groups had the same number of male and female participants.

The complex dynamic characteristics of the three groups of heartbeats are consistent with the complexity losing theory[32]: the heartbeat regulation system of healthy young people is highly complex, and its complexity decreases with decreasing aging heartbeat regulation function. Hence, heart failure will lead to less complex heartbeat characteristics.

To determine the fpTIR in CHF and healthy elderly and young heartbeat signals, the dimension $m$ of fuzzy permutation is set to 2 and 3, and the delay $\tau$ is set to 1 and 2, respectively; the membership $\alpha = 0.1–1$ (step size 0.1). As shown in Fig. 3, the fpTIR and pTIR behaved according to the complexity losing theory, and the fpTIR can more accurately characterize the non-equilibrium heartbeats. When $m = 2$ and $\tau = 1$, the discriminative ability of the fpTIR was higher than that of the pTIR. When $m = 3$ and $\tau = 1$, the pTIR of the CHF heartbeats is abnormally higher than that of healthy elderly people, whereas fpTIR correctly reflects the change in the complex characteristics of heartbeats, that is, healthy young > healthy elderly > CHF heartbeats. The fpTIR of the heartbeats was not significantly affected by the membership parameter $\alpha$, especially when $m = 3$; the results also do not change significantly in the control parameter interval of [0.1,1.0].

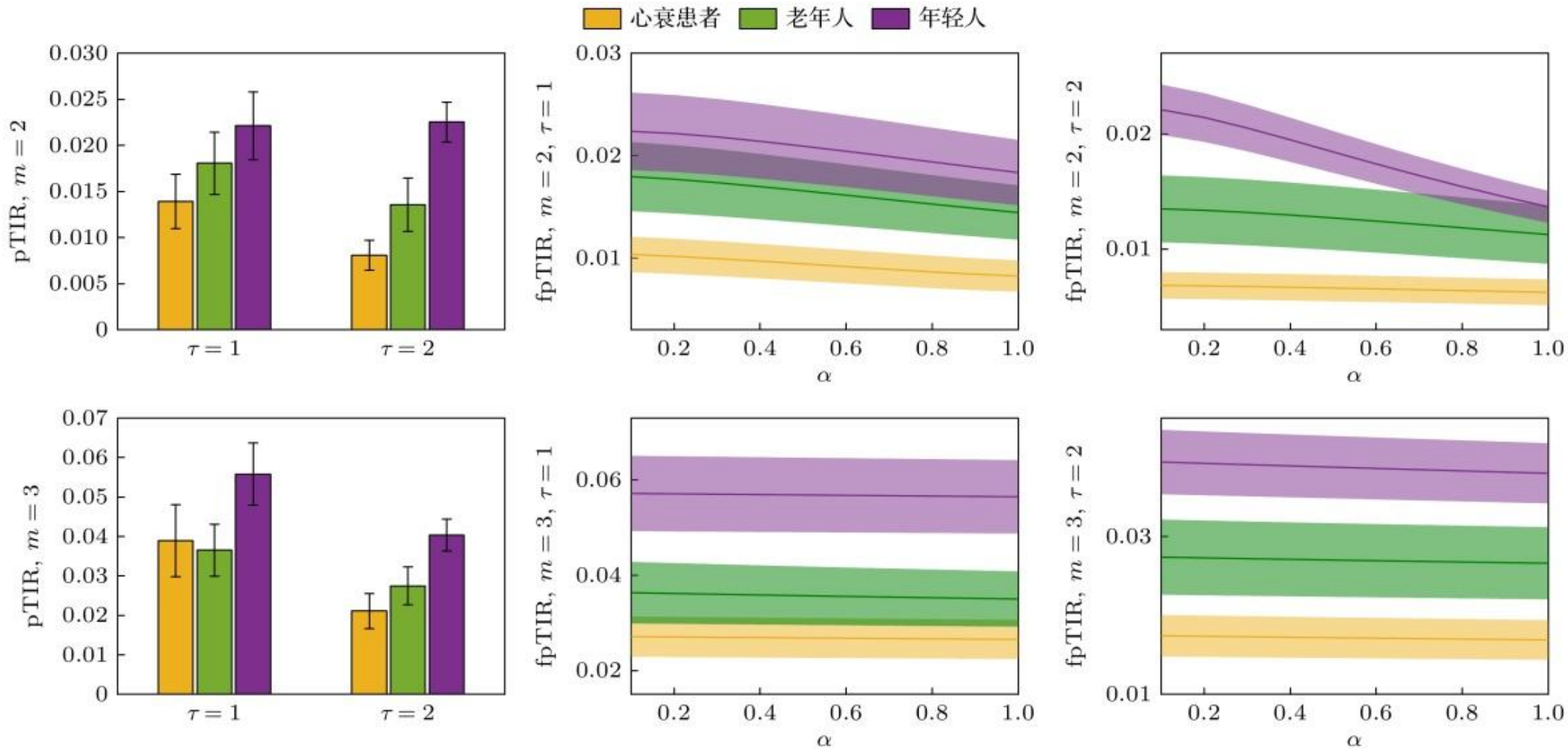

**Figure 3. The fpTIR and pTIR (mean ± standard error) of the heartbeats of CHF patients and healthy elderly and young participants. The dimension and delay of amplitude permutations are set to $m = 2$ and 3 and $\tau = 1$ and 2; and $\alpha = 0.1–1$ with step size of 0.1.**

According to the results of the heartbeat fpTIR analysis, the permutation membership $\alpha$ is set to 0.1. Combined with the probability distributions of fuzzy permutations, the fpTIR and fPEn of the three groups are compared, as shown in Fig. 4.

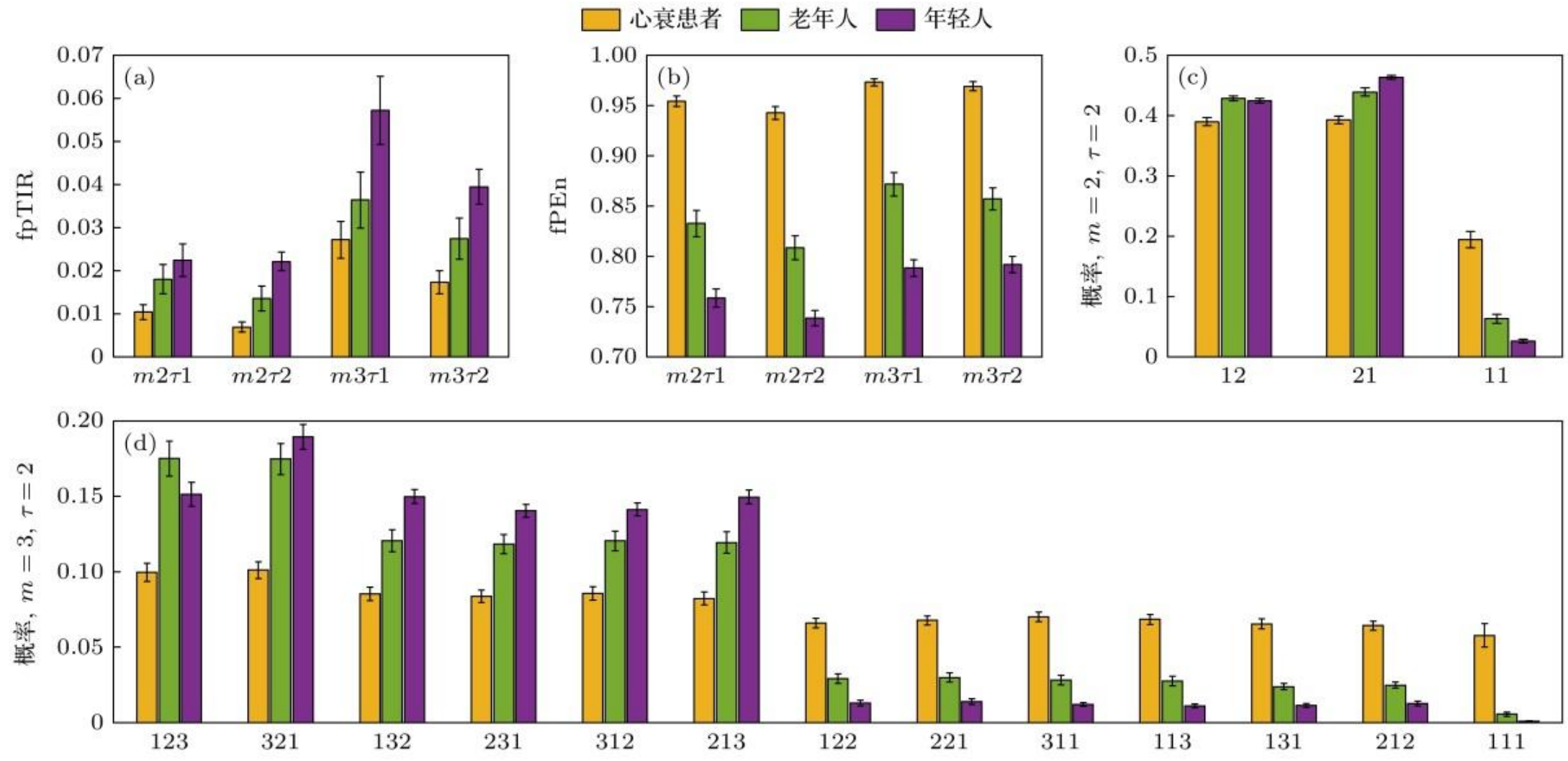

**Figure 4. The fpTIR, pTIR, and probabilities of permutations (mean ± standard error) of the heartbeats of CHF patients and healthy elderly and young participants. The membership degree of amplitude permutations $\alpha$ was set to 0.1. In (a) and (b) $m2\tau1$ represents dimension of 2 and delay of 1, and so on; and (c), (d) dimension and delay of amplitude permutations are $m$ = 2, $\tau$ = 2 and $m$ = 3, $\tau$ = 2.**

According to the Fig. 4(a) and 4(b), the fpTIR and fPEn of the three groups of heartbeats showed opposite trends. In particular, the fpTIR is consistent with the complexity losing theory, whereas the fPEn has contradictory results. Similar to the discrepancy in the surrogate data results, this discrepancy can be attributed to the different ways of analyzing the permutation probability. When $m$= 2, there are three permutations: up (12), down (21), and equality (11), in which equality is time reversible. The fpTIR is the difference between the probability of up and down, and the fPEn is the average amount of information contained in the three permutations. As shown in Fig. 4(c), CHF heartbeats had the smallest up-down probability difference, that is, the smallest fpTIR, smallest difference in the probability of the three permutations, and largest fPEn. Correspondingly, the healthy young group had the largest up-down probability difference and the largest probability difference of the three permutations; thus, they had the largest fpTIR and smallest fPEn. When $m = 3$, there are 13 amplitude permutations, of which self-symmetric 131, 212, and 111 mean time reversibility. As shown in Fig. 4(d), the probability distribution of fuzzy permutation of the three groups is relatively complex but consistent with that of $m = 2$. The probability difference of the symmetrical permutation of CHF heartbeats and that of all permutations are the smallest; thus, it has the smallest fpTIR and largest fPEn; by contrast, the healthy young group has the largest fpTIR and smallest fPEn. Therefore, the fpTIR accurately represented the complexity loss theory related to heartbeat signals, which is consistent with the physiological and pathological characteristics of heartbeat regulation. However, the fPEn has a better discrimination power among the three groups and is not sensitive to dimension and delay values. Therefore, the fpTIR and fPEn, as well as their relationship, requires further study.

Through model data validation and heartbeats analysis, the fpTIR can effectively characterize the characteristics of complex systems based on the absolute difference between vector elements to optimize the permutation, thus improving the accuracy of non-equilibrium analysis of complex systems.

## 4. Discussion

The relationship of fpTIR and fPEn in complex system analysis requires further analysis. The fpTIR is the probability difference[1,4–8] of the forward and backward permutations sequence or symmetric permutations, and the fPEn is the average information content[3,26,33] of the probability distribution of all permutations, which is the fundamental reason for their different results. Therefore, the smaller the difference in the permutation probability distribution, the higher the Shannon entropy, and the lower the TIR. The fuzzy permutation probability distribution of PhysioNet heartbeats[30,31] highlights the difference and even the opposite results between fpTIR and fPEn. Based on the numerical analysis, there should be a more complex relationship between fpTIR and fPEn. In the extreme case, if the probability distribution of all symmetric permutations is the same, the fpTIR is 0, and the fPEn changes with the probability distribution of permutations; in another extreme case, if all permutations are single permutations (the symmetric permutation of single permutations or the corresponding permutation in the reverse order is forbidden permutation)[1,7–9,15], then the fpTIR is 1, and the fPEn still varies with the difference of the probability distribution of permutations. The differences in the time irreversibility and entropy results are consistent with the results of previous studies (such as sleep EEG[1] and heartbeat[16]). Thus, for complex systems, a joint analysis of time irreversibility and entropy can explore their characteristics more comprehensively.

Fuzzy permutation has higher accuracy and can extract the features of complex systems more accurately; however, this can be achieved at the cost of higher implementation complexity and lower anti-interference ability. Because establishing fuzzy permutation requires the calculation of the absolute difference of vectors, the membership degree needs to be introduced into the permutation, requiring additional requirements for algorithm programming and hardware design in signal processing. In addition, any changes in the signal will modify the membership degree; thus, fuzzy permutation analysis is more sensitive to interference and noise. By contrast, the traditional permutation analysis, which is a typical symbolic time-series analysis method, has the main advantages of easy implementation and anti-noise performance. Therefore, in real-world systems analysis, if the signal interference is small, the fuzzy permutation method can improve the accuracy; otherwise, if it is significant, the traditional permutation analysis method may be more appropriate.

## 5. Conclusion

Fuzzy permutation can accurately extract the structural characteristics of sequences, thus improving the time irreversibility analysis of complex systems. The study conclusions are as follows:

1) Fuzzy permutation effectively extracts the precise temporal structure of the vector, taking contribution of the permutation membership of vectors. According to the theory of surrogate data, the fpTIR can effectively analyze the complex characteristics. In the complex physiological signal analysis, it represents the complexity losing theory of heartbeat more accurately and has high stability. However, fuzzy permutation analysis has higher requirements for algorithm programming and hardware design and is more sensitive to noise.

2) There are differences between the fpTIR and fPEn in the analysis of model sequence and heartbeat signals. Entropy and time irreversibility are both physical concepts, which can be used to quantify complex systems through information theory. However, owing to their different methods of dealing with probability distribution, they may produce different or even opposite results. The correlation analysis of fpTIR and fPEn connects the non-equilibrium and entropy complexity, which provides a valuable reference for studying the characteristics of complex systems from a broader perspective.